\documentclass[prb,twocolumn,superscriptaddress,showpacs,unsortedaddress]{revtex4}
\usepackage{graphicx}
\usepackage{dcolumn}
\usepackage{amsmath}

\newcommand{\vect}[1]{\mathbf{#1}}
\newcommand{\un}[1]{\,\mathrm{#1}}
\newcommand{\chem}[1]{\mathrm{#1}}

\begin{document}

\title{Lock-in transitions in $\chem{ErNi_2B_2C}$ and $\chem{TbNi_2B_2C}$}

\author{C. Detlefs}
\email{detlefs@esrf.fr}

\affiliation{European Synchrotron Radiation Facility, Bo\^{\i}te
  Postale 220, 38043 Grenoble Cedex, France}

\author{C. Song} \thanks{Current address: Electron Spin Science Center,
  Department of Physics, Pohang University of Science \& Technology,
  Pohang, 790-784, South Korea.}

\affiliation{Ames Laboratory and Department of Physics and Astronomy, 
        Iowa State University, Ames, IA 50011}

\author{S. Brown}
\author{P. Thompson}

\affiliation{XMAS CRG, European Synchrotron Radiation Facility,
  Bo\^{\i}te Postale 220, 38043 Grenoble Cedex, France }

\author{A. Kreyssig}

\affiliation{Institut f{\"u}r Angewandte Physik (IAPD), Technische
  Universit{\"a}t Dresden, 01062 Dresden, Germany}

\author{S. L. Bud'ko}
\author{P. C. Canfield}

\affiliation{Ames Laboratory and Department of Physics and Astronomy, 
        Iowa State University, Ames, IA 50011}

\date{Version \today}

\pacs{74.25.Ha, 75.25.+z, 61.10.Eq}

\begin{abstract}  
  High resolution x-ray magnetic scattering has been used to determine
  the variation with temperature of the magnetic modulation vector,
  $\vect{\tau}$, in $\chem{ErNi_2B_2C}$ and $\chem{TbNi_2B_2C}$ to
  study the interplay between the weakly ferromagnetic (WFM) phase and
  proposed lock-in transitions in these materials. At temperatures
  below the WFM transitions, the modulation wave vectors are within
  the resolution limit of the commensurate values $11/20$ and $6/11$
  for $\chem{ErNi_2B_2C}$ and $\chem{TbNi_2B_2C}$, respectively.
\end{abstract}

\maketitle

Investigations of the physical properties of the superconducting
rare-earth nickel boride-carbides, $R\chem{Ni_2B_2C}$ ($R$ =
$\chem{Gd}$--$\chem{Lu}$, $\chem{Y}$), continue to provide
insight into the interplay between superconductivity and magnetism
(For recent overviews, see refs.~\onlinecite{Natobook,Mueller01}.)

Recently, $\chem{ErNi_2B_2C}$ attracted special attention when the
possible coexistence of superconductivity and weak ferromagnetism
(WFM) was indicated by several
measurements\cite{Canfield96,Kawano99,Gammel00}. A similar WFM state
also exists in $\chem{TbNi_2B_2C}$\cite{Cho96,Dervenagas96,Song01b}.
The similarity of the crystallographic and magnetic properties of
these compounds suggests a common origin of the WFM\cite{Walker03}. In
order to study the interplay between WFM and the dominant
antiferromagnetic (AFM) order we have performed a comparative study of
these two compounds using the technique of x-ray resonant magnetic
scattering.

In $\chem{ErNi_2B_2C}$, superconductivity is observed below $T_C =
10.5 \un{K}$.  Neutron diffraction experiments
\cite{Sinha95,Zarestky95} show that below $T_N=6.0\un{K}$ it orders in
a transverse spin density wave with modulation wave vector
$\vect{\tau}_a \approx (0.55, 0, 0)$ and magnetic moments aligned
parallel to the $(0,1,0)$-axis of the crystal. As the temperature is
lowered, higher harmonic satellites develop, indicating that the spin
density wave squares up; the modulation wave vector was reported to be
approximately independent of temperature\cite{Sinha95,Zarestky95}.

A phase transition into a state with a weak ferromagnetic (WFM)
component of about $0.33\un{\mu_B}/\chem{Er}$ (at $T=2\un{K}$) is
observed at
$T_{\mathrm{WFM}}\approx2.3\un{K}$\cite{Canfield96,Kawano99,Gammel00}.
Whereas this transition is clearly resolved in zero-field specific
heat measurements\cite{Canfield96}, the magnetic moment of the ground
state has to be extrapolated from magnetization measurements at finite
fields above $H_{C1}$\cite{Canfield96}.  The structure and origin of
this WFM state remain unclear. Recently, neutron
scattering\cite{Choi01,Kawano02} showed that the WFM state is
intimately linked to the appearance of \emph{even} order harmonic
components of the spin wave and a lock-in of the modulation wave
vector, $\vect{\tau}$, onto a commensurate position.  However, the
values of $\vect{\tau}$ these two groups cite do not agree: Whereas
\citeauthor{Choi01}\cite{Choi01} find $\vect{\tau}=0.548$,
\citeauthor{Kawano02}\cite{Kawano02} report $\vect{\tau}=11/20 =
0.55$. Finally, an increase of the scattered intensity at nuclear
Bragg peaks confirmed the presence of a WFM
component\cite{Kawano99,Choi01}.

Because of the striking similarities in their magnetic properties, it
is useful to compare $\chem{ErNi_2B_2C}$ to $\chem{TbNi_2B_2C}$.
Whereas the latter is not superconducting, it displays the same local
moment anisotropy in the paramagnetic phase (easy axes $(1,0,0)$ and
$(0,1,0)$) and forms an AFM structure closely related to that of
$\chem{ErNi_2B_2C}$. In neutron scattering
experiments\cite{Dervenagas96} the magnetic modulation wave vector is
found to decrease from $\vect{\tau} =(0.551, 0, 0)$ at
$T_N=14.9\un{K}$ to $(0.545, 0, 0)$ below
$T_{\mathrm{WFM}}\approx7\un{K}$. Unlike in the $\chem{Er}$
compound, the $\chem{Tb}$ magnetic moments are aligned
\emph{parallel} to the modulation wave vector, forming a
\emph{longitudinal} spin wave.  Furthermore, a phase transition to a
WFM state similar to that of $\chem{ErNi_2B_2C}$, albeit in the
absence of superconductivity, occurs around $T_{\mathrm{WFM}}\approx
7\un{K}$\cite{Cho96,Sanchez98}, with magnetic moment $\approx
0.55\un{\mu_{B}}/\chem{Tb}$ at $T=2\un{K}$. Again, the neutron
experiments\cite{Dervenagas96} indicate that the appearance of a WFM
moment may be related to a lock-in of the AFM spin wave to a
commensurate propagation vector, in this case $\tau = 6/11 = 0.545$.

A recent Ginzburg-Landau type analysis\cite{Walker03} shows that a
lock-in to $\vect{\tau}=M/N \vect{G}$ ($\vect{G}$ = reciprocal lattice
vector) with $M$ even and $N$ odd directly induces weak ferromagnetism
as secondary order. This result is confirmed by a mean field
analysis\cite{Jensen02} of $\chem{ErNi_2B_2C}$.

In both compounds the magnetic ordering transitions are accompanied by
structural distortions which lower the crystal symmetry from
tetragonal (space group $I4/mmm$) to orthorhombic
($Immm$)\cite{Detlefs97b,Song99,Kreyssig01}. This distortion breaks
the symmetry between the $(1, 0, 0)$ and $(0, 1, 0)$ crystallographic
directions and thus leads to an unique easy axis of the magnetic
moment ($\mu \parallel (0, 1, 0)$ in $\chem{ErNi_2B_2C}$, and $\mu
\parallel (1, 0, 0)$ in $\chem{TbNi_2B_2C}$) and the modulation wave
vector ($\tau \parallel (1, 0, 0)$ in both
compounds\cite{Detlefs99,Song01}). Since these magneto-elastic effects
were not reported in the earlier
studies\cite{Dervenagas96,Kawano02,Choi01}, it appears doubtful that
they had the resolution necessary to clearly resolve a lock-in
transition into a commensurate state. We therefore performed
high-resolution resonant magnetic scattering experiments using
synchrotron x-rays\cite{Gibbs88,Hannon88} on both
$\chem{ErNi_2B_2C}$ and $\chem{TbNi_2B_2C}$.

\begin{figure}
  \includegraphics[width=0.95\columnwidth]{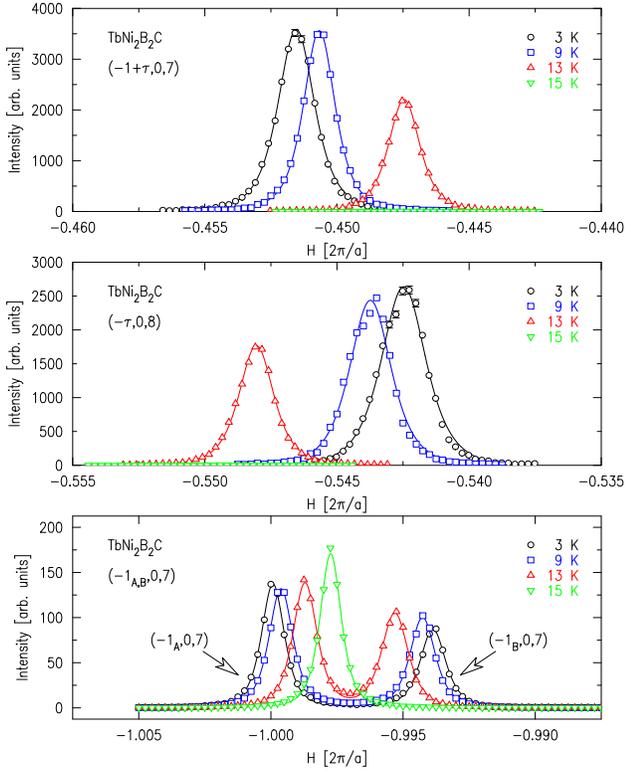}
  \caption[]{\label{fig.rawdata} Scans along the $(H,0,0)$ direction of
    reciprocal space through the $(-1+\tau,0,7)$ (top), $(-\tau,0,8)$
    (middle), and $(1_{A,B},0,7)$ (bottom) reflections of
    $\chem{TbNi_2B_2C}$ at selected temperatures. The horizontal
    scale is given by the experimental orientation matrix defined at
    $T=1.7\un{K}$. Note that the shift in position of the magnetic
    peaks is comparable to the shift of the structural peak due to the
    magneto-elastic effects. The width of the magnetic peaks is
    approximately $1.6\times 10^{-3} \frac{2\pi}{a}$.}
\end{figure}

\begin{figure}
  \centerline{%
    \includegraphics[width=0.95\columnwidth]{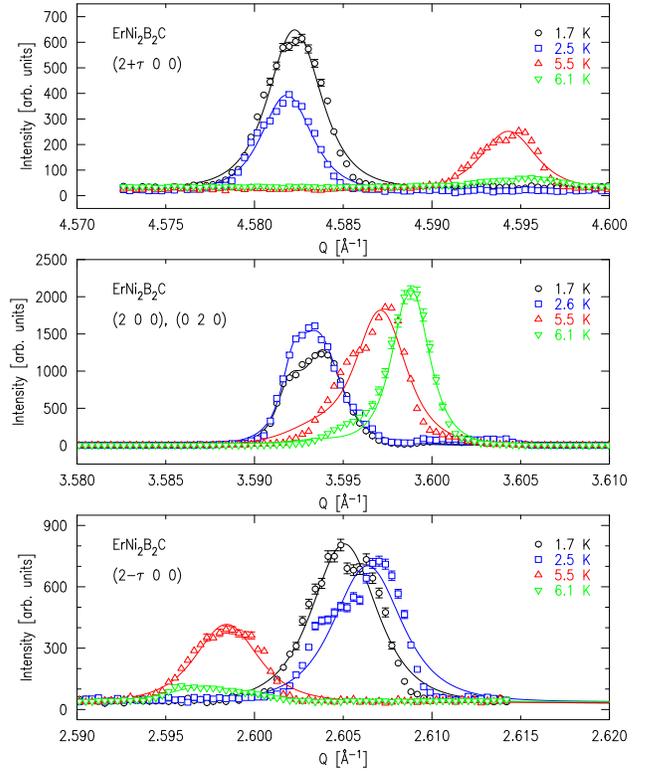}
  }
  \caption[]{\label{fig.er.rawdata} Longitudinal scans through 
    the $(2\pm\tau,0,0)$ (top, bottom) and $(2,0,0)$ (middle)
    reflections of $\chem{ErNi_2B_2C}$ at selected temperatures. The
    width of the peaks is approximately $2.0\times
    10^{-3}\frac{2\pi}{a}=3.6\times 10^{-3}\un{\AA^{-1}}$.}
\end{figure}

\begin{figure*}
  \parbox[t]{0.45\textwidth}{%
    \hspace*{\fill}
    \includegraphics[scale=0.55]{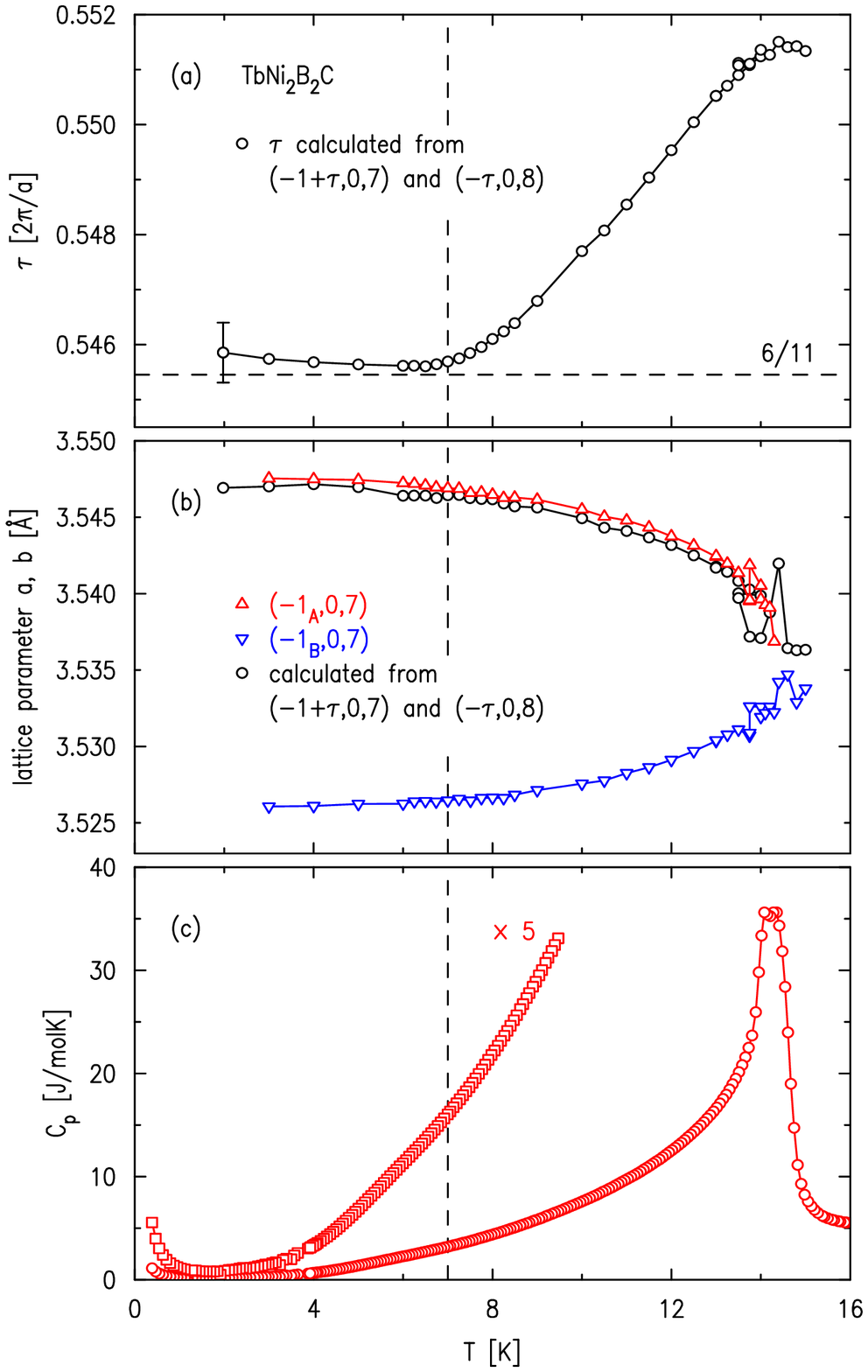}
    \caption[]{\label{fig.tau} (a) Around $7\un{K}$, the modulation
      wave vector, $\tau$, of $\chem{TbNi_2B_2C}$ locks in to the
      commensurate value of $6/11$, indicated by the dashed line.  (b)
      Basal plane lattice parameters, $a$ and $b$, extracted from the
      position of the $(-1_{A,B},0,7)$ reflections (triangles), and
      from the position of the $(-1+\tau,0,7)$ and $(-\tau,0,8)$
      magnetic reflections (circles). (c) The specific heat shows a
      clear $\lambda$-shaped maximum at the N{\'e}el transition. There
      is no signature of the lock-in transition. }  }
  \hspace*{0.003\textwidth}
  \parbox[t]{0.45\textwidth}{%
    \includegraphics[scale=0.55]{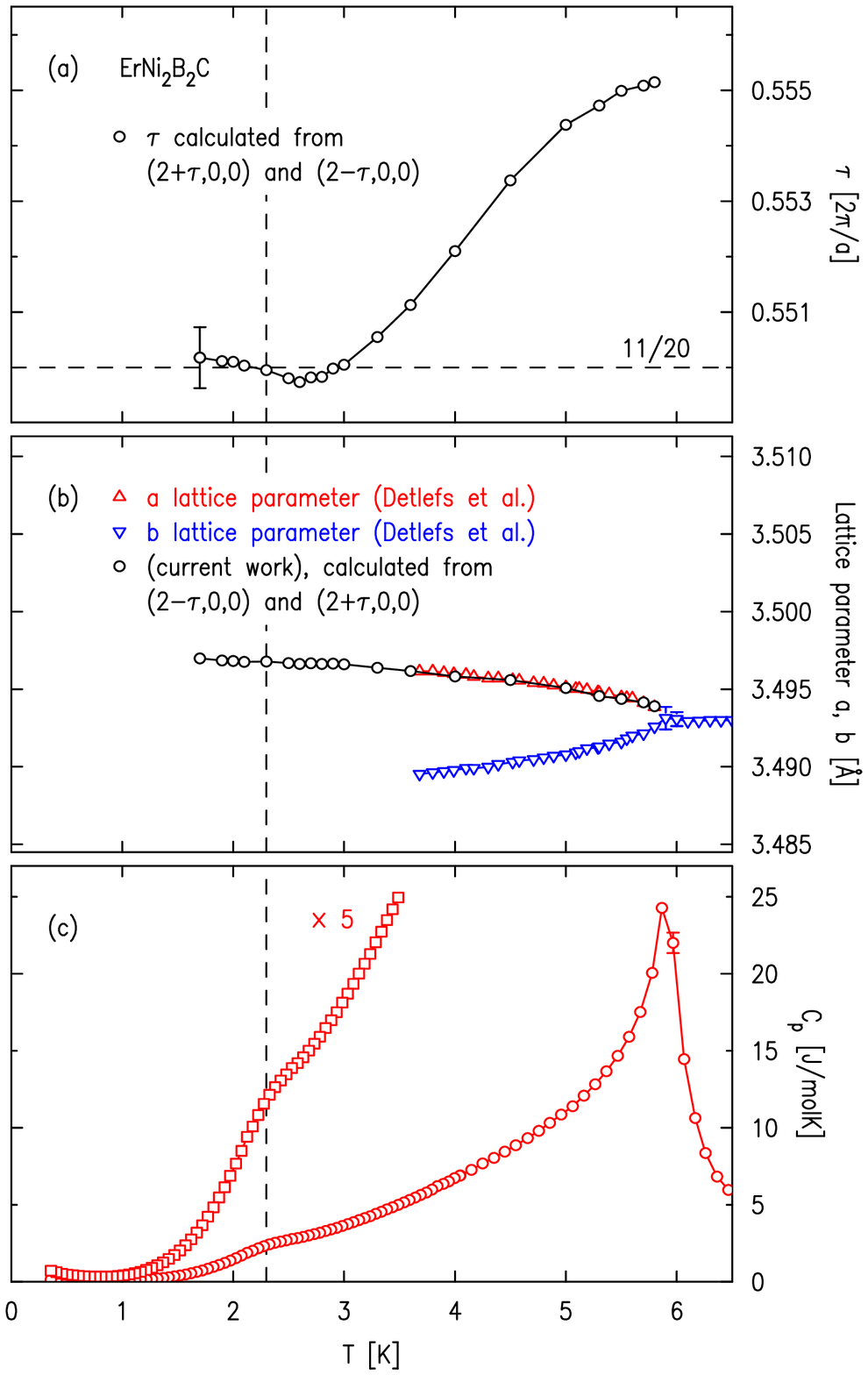}
    \hspace*{\fill}
    \caption[]{\label{fig.er.tau} (a) Around $3\un{K}$, the modulation
      wave vector, $\tau$, of $\chem{ErNi_2B_2C}$ locks in to the
      commensurate value of $11/20$, indicated by the dashed line.
      (b) Basal plane lattice parameters, $a$ and $b$, taken from
      Ref.~\onlinecite{Detlefs97b} (triangles), and from the position
      of the $(2+\tau,0,0)$ and $(2-\tau,0,0)$ magnetic reflections
      (circles). (c) In addition to the $\lambda$-shaped maximum at
      the N{\'e}el transition, a broad anomaly is observed near the
      WFM transition.}  }
\end{figure*}

Single crystals of $\chem{ErNi_2B_2C}$ and $\chem{TbNi_2B_2C}$ were
grown at the Ames Laboratory using a high-temperature flux growth
technique\cite{Canfield94,Cho95b}. Platelets extracted from the flux
were examined by x-ray diffraction and were found to be high quality
single crystals with the $(0, 0, 1)$-axis perpendicular to their flat
surface. The $\chem{ErNi_2B_2C}$ sample was cut perpendicular to the
$(1, 0, 0)$ direction and the resulting face was mechanically polished
to obtain a flat, oriented surface for x-ray diffraction. The sample
dimensions after polishing were approximately $2 \times 1.5 \times 0.5
\un{mm^3}$. The $\chem{TbNi_2B_2C}$ sample, the same one used in our
earlier studies\cite{Song99,Song01}, had a $(0, 0, 1)$ polished face.

The synchrotron experiments were carried out at XMAS CRG and at the
Tro{\"\i}ka undulator beamline (ID10C) of the European Synchrotron
Radiation Facility (ESRF). The samples were mounted on the cold finger
of a Displex closed cycle refrigerator equipped with an additional
Joule-Thompson stage\footnote{This cryostat is a prototype of a system
now commercially available from A.S. Scientific Products Ltd,
Abingdon, UK.}. The base temperature of this configuration was
approximately $1.7\un{K}$.

Fig.~\ref{fig.rawdata} shows scans through selected magnetic and
charge reflections of $\chem{TbNi_2B_2C}$.  The variation of the
$a$-axis lattice parameter is significant and has to be taken into
account when calculating the modulation wave vector\cite{Detlefs99}
(the horizontal scale is identical in all three panels).  The
corresponding raw data for $\chem{ErNi_2B_2C}$ are shown in
Fig.~\ref{fig.er.rawdata}.

The data presented in Fig.~\ref{fig.rawdata} were fitted to a
Lorentzian-squared line shape. $\tau$ in units of the reciprocal
lattice was calculated from
\begin{eqnarray}
        \frac{2\pi}{a} 
& = &   - Q_H(-1 + \tau, 0, 7) - Q_H(-\tau, 0, 8) 
\\
        \tau           
& = &   \frac{
                Q_H(-\tau, 0, 8)
        }{
                Q_H(-1 + \tau, 0, 7) + Q_H(-\tau, 0, 8)
        },
\end{eqnarray}
where $\vect{Q}=(Q_H,Q_K,Q_L)$ is the scattering vector. The resulting
data are presented in Fig.~\ref{fig.tau}.

For the case of $\chem{ErNi_2B_2C}$ a similar calculation was applied
to the measured positions of the $(2 \pm \tau, 0, 0)$ magnetic Bragg
reflections\cite{Detlefs99}:
\begin{eqnarray}
  \frac{2\pi}{a} 
  & = & 
  \frac{1}{2}\left[Q(2+\tau,0,0) + Q(2-\tau,0,0)\right]
  \\
  \tau 
  & = & 
  \frac{%
    \left[Q(2+\tau,0,0) - Q(2-\tau,0,0)\right]
    }{%
    \left[Q(2+\tau,0,0) + Q(2-\tau,0,0)\right]
    }.
\end{eqnarray}
The resulting data are presented in Fig.~\ref{fig.er.tau}.  

For both samples, the directly measured lattice parameter is in good
agreement with the one calculated from the position of the magnetic
reflections and with our earlier experiments
\cite{Detlefs97,Detlefs99,Song99,Song01}. We estimate the relative
systematic errors to be below $1\times 10^{-3}$, as indicated in
Figs.~\ref{fig.tau} and~\ref{fig.er.tau}. Clearly, these are
systematic errors -- the noise level is at least a factor of 10
smaller.

The importance of using this procedure is underlined by the
discrepancy between our value and that given by \citeauthor{Choi01}
\cite{Choi01}.  The latter, $\tau_{\mathrm{Choi}}=0.548$, can be
reproduced from our raw data by simply dividing the low temperature
$Q$-value of the $(2+\tau,0,0)$ reflection, $Q=4.5823\un{\AA^{-1}}$ by
the reciprocal lattice parameter at $T=6.1\un{K}$,
$\frac{2\pi}{a}=\frac{1}{2}\cdot Q(2,0,0)=\frac{1}{2}\cdot %
3.5988\un{\AA^{-1}}=1.7994\un{\AA^{-1}}$.  Using the high temperature
value of $Q(2,0,0)$ corresponds to taking the average of the
$Q$-values of the $(2,0,0)$ and $(0,2,0)$ peaks. We obtain
$Q/\frac{2\pi}{a} = 2.547 \approx 2 + \tau_{\mathrm{Choi}}$.
  
In both compounds the variation of $\tau$ vs.{\ }temperature flattens
out dramatically as $T_{\mathrm{WFM}}$ is approached; within the error
cited above their modulation wave vectors agree with the commensurate
values indicated by the dashed lines in Figs.~\ref{fig.tau}
and~\ref{fig.er.tau}.  Their absolute values, $\tau_{\chem{Er}} =
11/20$, and $\tau_{\chem{Tb}} = 6/11$, probably differ because of
small difference in the lattice parameters and electronic structures,
resulting in different spacings between the nested parts of the Fermi
surfaces. 

However, there is no sign of a discontinuous change of the modulation
at the proposed WFM transition. On the contrary, it appears as if the
modulation wave vector was changing continuously, and as if it was not
constant, i.e.{\ }not locked to a commensurate value, below
$T_{\mathrm{WFM}}$.  This behavior is confirmed by the neutron
data\cite{Choi01,Kreyssig03}.

Our results are consistent with a second order lock-in transition to
the commensurate values given above\cite{Walker03}. In this scenario,
the magnetic structure is composed of commensurate blocks separated by
domain walls often called ``discommensurations''\cite{Walker03}.
Across these domain walls the phase of the spin density wave shifts by
a defined value, in a way similar to the ``spin slip'' structures
observed in $\chem{Ho}$\cite{Gibbs85}. The modulation wave vector
$\tau$ of the structure is obtained from the Fourier transform over
several blocks, including the domain walls. The discrepancy between
the actual $\tau$ and the commensurate value is then proportional to
the average phase shift per unit length along the modulation wave
vector, i.e.{\ }the density of domain walls. When the temperature of
the system is varied the domain walls shift until in the limit $T
\rightarrow 0$ the distance between them becomes infinite and the wave
vector truly commensurate.  For $\chem{ErNi_2B_2C}$ and
$\chem{TbNi_2B_2C}$, this theory has to be modified to take into
account domain wall pinning which is expected to be significant
because of the large magneto-elastic strain induced by the
orthorhombic distortion associated with the AFM phase
transition\cite{Detlefs97b,Song99,Kreyssig01}.

The net ferromagnetic moment per formula unit resulting from the above
commensurate structures would be $2/N$ of the saturation moment, i.e.,
$1/10\un{\mu_\chem{Er}}$ and $2/11\un{\mu_\chem{Tb}}$,
respectively. These numbers have to be compared to the measured
fractions (at $T=2\un{K}$),
$0.33\un{\mu_{B}}/7.8\un{\mu_{B}}\approx1/24$ for
$\chem{ErNi_2B_2C}$\cite{Canfield96}, and
$0.55\un{\mu_{B}}/9.5\un{\mu_B}\approx 1/17$ for
$\chem{TbNi_2B_2C}$\cite{Cho96}. That the observed WFM moment is
significantly smaller than the predicted value might indicate that the
saturated, completely ordered state was not reached at the measurement
temperature, or that only a finite volume fraction of the sample
undergoes the WFM transition. Furthermore, in $\chem{ErNi_2B_2C}$ the
observed magnetization is lowered by the diamagnetism of the
superconducting state. Note also that $\tau_{\chem{Er}} = 11/20$ does
not have the form $N/M\vec{G}$ with $N$ even and $M$ odd, so that the
lock-in does not necessarily induce weak
ferromagnetism\cite{Walker03}.

Finally, we note that the specific heat of $\chem{ErNi_2B_2C}$
exhibits a broad anomaly near the WFM transition
(Fig.~\ref{fig.er.tau}(c)), whereas the specific heat of
$\chem{TbNi_2B_2C}$ (Fig.~\ref{fig.tau}(c)) shows no indication of the
WFM transition. This might be related to the different form of the
commensurate value, odd/even vs.{\ }even/odd, for the $\chem{Er}$ and
$\chem{Tb}$ compound, respectively \cite{Walker03}.

In summary, we have performed high resolution magnetic x-ray
diffraction measurements to study the AFM host structures of
$\chem{ErNi_2B_2C}$ and $\chem{TbNi_2B_2C}$ close to the weak
ferromagnetic transition at $T_{\mathrm{WFM}}$. We determined the
modulation wave vector $\tau$ with high precision, taking into
consideration the magneto-elastic distortion associated with the
antiferromagnetic order. In both materials above $T_{\mathrm{WFM}}$
$\tau$ varies strongly with temperature, whereas below
$T_{\mathrm{WFM}}$ the variations are much smaller, but still finite
and observable. Across the whole temperature range the variations are
continuous, so that first order transitions can be excluded. Our
observations are consistent with a second order lock-in transition in
the presence of discommensurations\cite{Walker03}.

\acknowledgments

The authors are grateful to A.~I.~Goldman, H.~Furukawa-Kawano,
P.~L.~Gammel, and M.~B.~Walker for stimulating discussions. We also
wish to thank the ESRF and the beamline staff of ID10C for assistance
with the experiments. Ames Laboratory is operated for the
U.~S.~Department of Energy by Iowa State University under Contract
No.~W-7405-Eng-82. This work was supported by the Director for Energy
Research, Office of Basic Sciences.


\end{document}